\documentclass[onecolumn,amsmath,amssymb,11pt]{revtex4}
\usepackage{psfrag}
\usepackage{epsfig}
\usepackage{graphicx,graphics}% Include figure files
 \normalsize

\def\lsim{\raise0.3ex\hbox{$\;<$\kern-0.75em\raise-1.1ex\hbox{$\sim\;$}}}
\def\gsim{\raise0.3ex\hbox{$\;>$\kern-0.75em\raise-1.1ex\hbox{$\sim\;$}}}

\def\2tvec#1#2{ \left( \begin{array}{c}
#1  \\
#2  \\
\end{array} \right)}%
\def\mat2#1#2#3#4{ \left( \begin{array}{cc}
#1 & #2 \\
#3 & #4 \\
\end{array} \right) }%
\def\Mat3#1#2#3#4#5#6#7#8#9{ \left( \begin{array}{ccc}tri-bimaximal
#1 & #2 & #3 \\
#4 & #5 & #6 \\
#7 & #8 & #9 \\
\end{array} \right) } %
\def\Mat3#1#2#3#4#5#6#7#8#9{ \left(
\begin{array}{ccc}
#1 & #2 & #3 \\
#4 & #5 & #6 \\
#7 & #8 & #9 \\
\end{array} \right) }

\def\3tvec#1#2#3{ \left( \begin{array}{c}
#1  \\
#2  \\
#3  \\
\end{array} \right)}

\def\4tvec#1#2#3#4{ \left( \begin{array}{c}
#1  \\
#2  \\
#3  \\
#4  \\
\end{array} \right)}

\def\hbar{\hspace{1mm}\bar{}\hspace{-1mm}h}

  \def\m{\mu}

\newcommand{\balg}{\begin{align}}
\def\bea{\begin{eqnarray}}
\def\eea{\end{eqnarray}} \newcommand{\be}{\begin{eqnarray}}
\newcommand{\ee}{\end{eqnarray}}
\usepackage{feynmp} \usepackage{slashed} \usepackage{latexsym}
\usepackage{graphicx} \usepackage{verbatim}
\usepackage[normalem]{ulem} %%
\begin{document}

%%%%%%%%%%%%%%%%%%%%%%%%%%%%%%%%%%%%%%%%%%%%%%%%%%%%%%%%%%%%%%%%%%%%%%%%%%%%%%%%%%
\title{%\vspace{-2cm}
%\vglue -0.3cm
%\vskip 0.5cm
%\Large \bf
Fermion masses and flavor mixing in modular $A_4$ Symmetry}
\author{
{\textbf{Mohammed Abbas}}\thanks{email: \tt
maabbas@ju.edu.sa}
 \\
%%%%%%%%%%%%%%%%%%%%%%%%%%%%%%%%%%%%%%%%%%%%%%%%%%%%%%%%%%%%%%%%%
{\normalsize\em
Physics Department, Faculty of Science and Arts-Tabarjal, Jouf University} \\
{\normalsize\em Al-Jouf, KSA.
\vspace*{0.15cm}}
\\
{\normalsize\em
Physics Department, Faculty of Sciences, Ain Shams University,
} \\
{\normalsize\em Abbassiyah 11566, Cairo, Egypt.
 \vspace*{0.15cm}}
\\
}

\maketitle
\vspace{-0.8cm}

%%%%%%%%%%%%%%%%%%%%%%%%%%%%%%%%%%%%%%%%%%%%%%%%%%%%%%%%%%%%%%%%%%%%%%%%%%%%%%%%%%%
%%\begin{center}
%%{\Large\bf Is the tri-bimaximal mixing accidental? \\
%%\vskip .3cm}
%%\end{center}

%\textbf{Abstract}\\
%\begin{abstract}
\noindent
We consider a flavor model based on $A_4$ modular group to account for both lepton and quark parameters (masses and mixing). The inverse seesaw mechanism is considered to produce the light neutrino masses. Lepton masses and mixing are obtained in terms of Yukawa coupling ratios and values of the modulus $\tau$ nearby some fixed points for inverted neutrino mass hierarchy. The quark masses and mixing are arisen at the same $\tau$ values used in inverted neutrino mass hierarchy and are in agreement with the recent data.
%\end{abstract}

%\end{abstract}

\section{Introduction}
The smallness of the neutrino mass is usually explained via type-I seesaw mechanism \cite{T1} which is the common scenario to account for neutrino masses and mixing. In this mechanism, the small neutrino mass is obtained by the extension of the fermion content of the Standard Model (SM) with three Gauge singlets as heavy right handed neutrinos $N_i$. The mass of the light neutrino can be calculated through the relation $m_{\nu} = - m_D {M_R}^{-1} m_D^T$, where $m_D$ is the Dirac mass and $M_R$ is the Majorana mass of right handed neutrinos $N_i$. To account for the tiny neutrino mass of order ${\cal O}(10^{-2})$ eV, either the mass scale of $N_i$ will be of order ${\cal O}(10^{11}$ GeV) or one should consider a very small Dirac coupling for TeV mass scale of right handed neutrinos. The large right handed neutrino masses are far from experimental reach. In addition, the lepton number is violated via the large scale of the right handed neutrino masses.

On the other hand, the inverse seesaw mechanism \cite{Wyler:1982dd, Mohapatra:1986bd,
Ma:1987zm} is an alternative mechanism to account for the small neutrino mass by considering a small scale $\mu_s$ and making a double suppression by the
new scale $M_R$ via the relation $m_{\nu} =m_D {M_R}^{-1} \mu_s {M_R^T}^{-1}m_D^T$.
In the case of type-I seesaw, the lepton number violation (LNV) takes place
via the majorana mass term of $N_i$,
which is so large. Conversely, in the inverse seesaw,
the lepton number is violated by the very small mass $\mu_s$ of the singlet $S$ which is a very small scale
compared to the electroweak scale.
One can consider the lepton number as an approximate symmetry of nature, so it
is convenient to break it by a small amount
instead of a large mass like $M_R$.
According to 't Hooft \cite{'t Hooft}, if $\mu_s$ tends to zero, the neutrino mass $m_{\nu}$ goes to zero
and LNV vanishes so that the symmetry is enhanced.

The flavor symmetry was proposed to account for many aspects such as the differences in mixing and mass hierarchy for lepton and quark sectors. Several models based on discrete symmetries were proposed to account for fermion masses and mixing (see \cite{discrete-symmetries}). Most of these models suffer from considerations of a large number of scalars (flavons), considering extra $Z_N$ symmetries and/or fine tuning to account for experimental data.

Recently, finite modular groups $\Gamma_N$ have been proposed to explain the flavor aspects \cite{modulargroups, Feruglio:2017spp}. In such groups, the group transformations are extended to include the coupling constants which can transform non trivially. Extra symmetries under modular weights are impeded into the group, so there is no need to impose other symmetries to match the data. Some of $\Gamma_N$ are isomorphic to finite permutation groups, for instance, $\Gamma_2\cong S_3$ \cite{Kobayashi:2018vbk, Okada:2019xqk, Du:2020ylx, Xing:2019edp}, $\Gamma_3\cong A_4$ \cite{Kobayashi:2018scp, Okada:2018yrn, Ding:2019zxk, Gui-JunDing:2019wap, Nomura:2019yft, Nomura:2019xsb, Asaka:2019vev}, $\Gamma_4\cong S_4$ \cite{Penedo:2018nmg, Kobayashi:2019mna, Okada:2019lzv, Kobayashi:2019xvz, Wang:2019ovr} and $\Gamma_5\cong A_5$ \cite{Novichkov:2018nkm, Ding:2019xna, Criado:2019tzk}. Attempts have been made to account for both leptons and quarks using a modular group with a single modulus value for both leptons and quarks \cite{Okada:2020rjb, Okada:2019uoy, Liu:2020akv, Kobayashi:2019rzp, Kobayashi:2018wkl, Yao:2020qyy, Lu:2019vgm}. Models with different moduli for charged lepton and neutrino sectors have been studied using the concept of modular residual symmetries \cite{Novichkov:2018yse, Novichkov:2018ovf}. Multiple modular symmetries with more than one modular group have been discussed in \cite{deMedeirosVarzielas:2019cyj}. The double covering of modular groups has been investigated in \cite{Yao:2020zml, Wang:2020lxk, Novichkov:2020eep, Liu:2019khw}. The modular invariance combining with the generalized CP symmetry has been studied to predict CP violating phases of quarks and leptons \cite{Okada:2020brs, Novichkov:2019sqv, Kobayashi:2019uyt}. Most of the above models used either type-I seesaw mechanism or the non-renormalizable five dimensional operator to generate neutrino masses. The inverse seesaw mechanism has been used for some modular invariance models \cite{Nomura:2019xsb, Nomura:2020cog}.

In this paper, we introduce a model based on modular $A_4$ symmetry to account for masses and mixing for leptons and quarks. First, we give an introduction to the modular groups and how to use them as flavor symmetries, then we explain our $A_4$ model in the lepton sector and finally we study the quark masses and mixing.
\section{Modular groups}
The modular group $\bar{\Gamma}$ is defined as linear fractional transformations on the upper half of the complex plan ${\cal H}$ and has the form \cite{Bruinier2008The, diamond2005first, Gunning1962, Feruglio:2017spp}
\bea
\gamma: \tau\rightarrow \gamma(\tau)= \frac{a\tau+b}{c\tau+d},\label{gamma}
\eea
 where $a, b, c, d \in Z, ad-bc=1$. The modular group $\bar{\Gamma}$ is isomorphic to the projective special linear group
\bea
PSL(2, Z)=SL(2, Z)/\{I, -I\},
\eea
 where
\bea
SL(2, Z)=\Big\{ \left(
                  \begin{array}{cc}
                    a & b \\
                    c & d \\
                  \end{array}
                \right)
, a,b, c, d \in Z, ad-bc=1\Big\}.
\eea
The generators of the group $\bar{\Gamma}$ are two matrices $S$ and $T$ where their action on the complex number $\tau$ is given by,
\bea%
S:\tau\rightarrow \frac{-1}{\tau}, ~~~~T: \tau\rightarrow \tau+1.
\eea
In the two by two representation, the two generators $S, ~T$ can be represented as
\bea
S=\left(
    \begin{array}{cc}
      0 & 1 \\
      -1 & 0 \\
    \end{array}
  \right), ~~~~T=\left(
                   \begin{array}{cc}
                     1 & 1 \\
                     0 & 1 \\
                   \end{array}
                 \right).
                 \eea
%Because $I$ and $-I$ are indistinguishable in $PSL(2, Z)$,
They should satisfy the conditions  $$
S^2=\textbf{1}, ~~~~~~(ST)^3=\textbf{1}. \label{st_condition}
$$
Define the infinite modular groups $\Gamma(N), N=1, 2, 3, ....$ as
\bea
\Gamma(N)=\Big\{\left(
                    \begin{array}{cc}
                      a & b \\
                      c & d \\
                    \end{array}
                  \right)\in SL(2, Z), \left(
                    \begin{array}{cc}
                      a & b \\
                      c & d \\
                    \end{array}
                  \right)=\left(
                    \begin{array}{cc}
                      1 & 0 \\
                      0 & 1 \\
                    \end{array}
                  \right) mod ~N\Big\}.\label{Gamma(N)}
\eea
For $N=1$,
\bea
\Gamma(1)=\Big\{\left(
                    \begin{array}{cc}
                      a & b \\
                      c & d \\
                    \end{array}
                  \right)\in SL(2, Z), \left(
                    \begin{array}{cc}
                      a & b \\
                      c & d \\
                    \end{array}
                  \right)=\left(
                    \begin{array}{cc}
                      1 & 0 \\
                      0 & 1 \\
                    \end{array}
                  \right) mod ~1\Big\}.
\eea
Since any integers can satisfy the conditions $a, d=1 ~~mod ~1$ and $b, c=0 ~~mod~ 1$, $\Gamma(1)\equiv SL(2, Z)$.
For $N=1, 2$, we define $\bar{\Gamma}(N)=\Gamma(N)/\{I, -I\}$ whereas for $N>2$, $\bar{\Gamma}(N)=\Gamma(N)$ because $-I \not\in \Gamma(N)$ for $N>2$. It is straightforward to notice that $\bar{\Gamma}(1)=PSL(2, Z)=\bar{\Gamma}$. The groups $\bar{\Gamma}$ and its subgroup $\bar{\Gamma}(N)$ are discrete but infinite, while the quotient modular group $\Gamma_N=\bar{\Gamma}/\bar{\Gamma}(N)$ is finite. The group $\Gamma_N$ is called the finite modular group and can be obtained by extending the conditions on the generators with the condition $T^N=\textbf{1}$. For some N, the finite modular group $\Gamma_N$ is isomorphic to a permutation group, for instance, $\Gamma_2\cong S_3$, $\Gamma_3\cong A_4$, $\Gamma_4\cong S_4$ and $\Gamma_5\cong A_5$.

The modular function $f(\tau)$ of weight $2k$ is a meromorphic function of the complex variable $\tau$ satisfies
\bea%
f(\gamma(\tau))=f(\frac{a\tau+b}{c\tau+d})=(c\tau+d)^{2k} f(\tau) ~~~~\forall \left(
                                                                   \begin{array}{cc}
                                                                     a & b \\
                                                                     c & d \\
                                                                   \end{array}
                                                                 \right)\in \Gamma(N),\label{modular function}
                                                                 \eea
                                                                 where the integer $k\geq 0$.
By using Eqs. (\ref{gamma}) and (\ref{Gamma(N)}), it is easy to calculate
\bea
\frac{d(\gamma(\tau))}{d\tau}=\frac{1}{(c\tau+d)^2}.
\eea
From Eq. (\ref{modular function}), one can get $$\frac{f(\gamma(\tau))}{f(\tau)}=\Big(\frac{d(\gamma(\tau))}{d\tau}\Big)^{-k}, ~~~f(\gamma(\tau))d(\gamma(\tau))^k=f(\tau)d\tau^k.$$
From the above equation, we conclude that the k-form $f(\tau)d\tau^k$ is invariant under $\Gamma(N)$. If the modular function is holomorphic everywhere, it is called "modular form" of weight $2k$. The modular forms of level $N$ and weight $2k$ form a linear space of finite dimension. In the basis at which the transformation of a set of modular forms $f_i(\tau)$ is described by a unitary representation $\rho(\gamma)$, one can get
\bea
f_i(\gamma(\tau))=(c\tau+d)^{2k} \rho_{ij}(\gamma) f_j(\tau), ~~\gamma \in \Gamma(N).
\eea
Consider the superpotential  $W(z,\phi)$ be written in terms of supermultiplets $\phi^I$, where $I$ refers to different sectors in the theory,
\bea
W(\tau, \phi)=\sum_{I}\sum_{n} Y_{I_1~I_2~...I_n}(\tau) \phi^{I_1}...\phi^{I_n}.
\eea
The supermultiplets $\phi^I$ transforms under $\Gamma_N$ in the representation $\rho(\gamma)$ as
\bea
\phi^{(I)}(\tau)\rightarrow\phi^{(I)}(\gamma(\tau))=(c\tau+d)^{-2k} \rho^{(I)}(\gamma) \phi^{(I)}(\tau).
\eea
The invariance of the superpotential $W(z, \phi)$ under the modular transformation requires $Y_{I_1~I_2~...I_n}(z)$ to be a modular form transforming in the representation
\bea
Y_{I_1~I_2~...I_n}(\gamma \tau)=(cz+d)^{2k_Y(n)} \rho(\gamma) Y_{I_1~I_2~...I_n}(\tau).
\eea
The modular invariance forces the condition \bea
k_Y=k_{I_1}+k_{I_2}+...+k_{I_n}.\label{K_rules}
\eea

\subsection{Modular forms of level 3}
The group $A_4$ has one triplet representation $\textbf{3}$ and 3 singlets $\textbf{1}, ~\textbf{1}^{\prime}$ and $\textbf{1}^{\prime\prime}$ and is generated by two elements $S$ and $T$ satisfying the conditions
\bea
S^2=T^3=(ST)^3=\textbf{1}.
\eea
The modular form of level 3 has the form
$$f_i(\gamma(\tau))=(c\tau+d)^{2k} \rho_{ij}(\gamma) f_j(\tau), ~~\gamma \in \Gamma(3).$$ The modular form of weight 2 and level 3 transforms as a triplet and is given by $Y_3^{(2)}=(y_1, y_2, y_3)$, \cite{Feruglio:2017spp} where :
\bea
y_1(\tau) &=& \frac{i}{2\pi}\left[ \frac{\eta'(\tau/3)}{\eta(\tau/3)}  +\frac{\eta'((\tau +1)/3)}{\eta((\tau+1)/3)}
+\frac{\eta'((\tau +2)/3)}{\eta((\tau+2)/3)} - \frac{27\eta'(3\tau)}{\eta(3\tau)}  \right], \nonumber \\
y_2(\tau) &=& \frac{-i}{\pi}\left[ \frac{\eta'(\tau/3)}{\eta(\tau/3)}  +\omega^2\frac{\eta'((\tau +1)/3)}{\eta((\tau+1)/3)}
+\omega \frac{\eta'((\tau +2)/3)}{\eta((\tau+2)/3)}  \right] ,\nonumber \\
y_3(\tau) &=& \frac{-i}{\pi}\left[ \frac{\eta'(\tau/3)}{\eta(\tau/3)}  +\omega\frac{\eta'((\tau +1)/3)}{\eta((\tau+1)/3)}
+\omega^2 \frac{\eta'((\tau +2)/3)}{\eta((\tau+2)/3)} \right]\,.\label{Y_forms}
\eea
where $\omega=e^{2i\pi/3}$ and the Dedekind eta-function $\eta(z)$ is defined as
\bea
\eta(\tau)=q^{1/24} \prod_{n =1}^\infty (1-q^n), \qquad  q=e^{2\pi i\tau}.\,
\eea
One can construct modular forms of higher weights using the multiplication rules of $A_4$ \cite{Feruglio:2017spp}. Modular forms of weight 4 are constructed via multiplication of two triplets of weight 2. Using $A_4$ multiplication rules of two triplets, one can get one triplet and three singlets all of weight 4 as
\bea
Y_3^{(4)}=\left(
          \begin{array}{c}
            y_1^2-y_2~y_3 \\
            y_3^2-y_2~y_1 \\
            y_2^2-y_1~y_3 \\
          \end{array}
        \right), ~Y_1^{(4)}=y_1^2+2y_2~y_3, ~ Y_2^{(4)}=y_3^2+2y_2~y_1,~Y_3^{(4)}=y_2^2+2y_1~y_3.
        \eea
The representations of the above singlets are $$Y_1^{(4)}\sim 1, ~~~~~Y_2^{(4)}\sim 1^{\prime}, ~~~~~~Y_3^{(4)}\sim 1^{\prime\prime}.$$ At all values of $\tau$, the condition $Y_3^{(4)}=0$ is satisfied.

We will use the basis where the generators of $A_4$ in triplet representation are
\bea
S=\frac{1}{3}\left(
               \begin{array}{ccc}
                 -1 & 2 & 2 \\
                 2 & -1 & 2 \\
                 2 & 2 & -1 \\
               \end{array}
             \right), ~~~~T=\left(
                              \begin{array}{ccc}
                                1 & 0 & 0 \\
                                0 & \omega & 0 \\
                                0 & 0 & \omega^2 \\
                              \end{array}
                            \right).
                            \eea

\section{$A_4$ modular invariance model}
The lepton content in the model is extended by adding a triplet of chiral supermutiplets $N^c$ as a right handed neutrino and three SM singlets $S_i$ to get the neutrino masses via the inverse seesaw mechanism. We add a gauge singlet scalar $\chi$ transforming trivially under $A_4$ to get the masses of the singlet fermions $N^c$ and $S$. Contrary to most flavor symmetric models, no more flavons are considered and no extra discrete symmetries are considered in our model. According to the modular invariance condition in Eq.(\ref{K_rules}), we chose the modular weights such that the following relations are satisfied:
\bea
k_L+k_{H_d}+k_E&=&2,\nonumber\\
k_L+k_{H_u}+k_N&=&2,\nonumber\\
2k_S+4k_{\chi}&=&0,\nonumber\\
k_S+k_N+k_{\chi}&=&0.
\eea
If we chose $k_L=3, ~k_{H_u}=0$, we can get the modular weights of other fields as shown in Table (\ref{assignment 1}).
\begin{table}
  \centering
  \begin{tabular}{|c|c|c|c|c|c|c|c|c|c|}
  \hline
  fields & L & $E^c_1$ &$E^c_2$ &$E^c_3$& $N^c$ & S &$ H_d$ &$ H_u$ & $\chi$ \\
   \hline
  $A_4$ & 3 & 1 & $1^{\prime\prime}$ &$1^{\prime}$ & 3 & 3 & 1 &1& 1 \\
   \hline
  $k_I$ & 3 & -1 & -1 &-1 & -1 & 2 & 0&0 & -1 \\
  \hline
\end{tabular}
  \caption{Assignment of flavors under $A_4$ and the modular weight $k_I$}\label{assignment 1}
\end{table}
The lepton modular $A_4$ invariant superpotential can be written as
\bea
w_l&=&\lambda_1 E_1^c H_d(L \otimes Y_3^{(2)} )_1 +\lambda_2 E_2^c H_d(L \otimes Y_3^{(2)})_1^{\prime}+\lambda_3 E_3^c H_d(L \otimes Y_3^{(2)})_1^{\prime\prime}+g_1~
((N^c H_u L)_{3S}Y_3^{(2)})_1\nonumber\\&+&g_2~((N^c H_u L)_{3A}Y_3^{(2)})_1+
 h~ (N^c \otimes S)_1~\chi+\frac{f}{\Lambda^3}(S\otimes S)_1~\chi^4,\label{superpot.}
\eea
where $\Lambda$ is the non-renormalizable scale and $g_1$ is the coupling constant of the term of the symmetric triplet arising from the product of the two triplets $L$ and  $Y$, while $g_2$ is the coupling of the antisymmetric triplet term. After spontaneous symmetry breaking, the scalar fields $H_u$, $H_d$ and $\chi$ acquire vevs namely $v_u$, $v_d$ and $v^\prime$ respectively, where $v^\prime\gg v_u, v_d$. We assume that $v^\prime$ satisfies the relation $\frac{v^\prime}{\Lambda}\sim {\cal O}(\lambda_c)$ where $\lambda_c=0.22$ is the Cabibbo angle.
%\subsection{Charged Lepton massas}
From Eq.(\ref{superpot.}), we can write the charged lepton mass matrix as
\bea
m_e&=&v_d \left(
                      \begin{array}{ccc}
                        \lambda_1 & 0 & 0 \\
                        0 & \lambda_2 & 0 \\
                        0 & 0 & \lambda_3 \\
                      \end{array}
                    \right)\times \left(
      \begin{array}{ccc}
        y_1 & y_3 & y_2 \\
        y_2 & y_1 & y_3 \\
        y_3 & y_2 & y_1 \\
      \end{array}\right).
      \eea
To deal only with the left-handed mixing, it is convenient to use the Hermitian matrix $M_e=m_e^{\dagger} ~m_e$ which can be diagonalized as $$M_e^{diag}=U_e^{\dagger}M_eU_e .$$
%\subsection{Neutrino masses}
The neutrino mass matrices are
      \bea
\mu_s&=&f v^{\prime}\lambda_c^3\left(
                    \begin{array}{ccc}
                      1 & 0 & 0 \\
                      0 & 0 & 1 \\
                      0 & 1 & 0 \\
                    \end{array}
                  \right),~~~~~~~ M_R=h v^{\prime}\left(
                    \begin{array}{ccc}
                     1 & 0 & 0 \\
                      0 & 0 & 1 \\
                      0 & 1 & 0 \\
                    \end{array}
                  \right),\nonumber \\m_D &=& v_u\left(
  \begin{array}{ccc}
    2 g_1 y_1 & (-g_1+g_2) y_3 & (-g_1-g_2) y_2 \\
    (-g_1-g_2) y_3 & 2 g_1 y_2 & (-g_1+g_2) y_1 \\
    (-g_1+g_2) y_2 & (-g_1-g_2) y_1 & 2 g_1 y_3 \\
  \end{array}\right).
\eea
%In general, the charged lepton mass matrix, $M_e$, is not Hermitian so it can be diagonalized by two unitary matrices as, $$M_e^{diag}=U_e^{L\dagger}M_e~U_e^R,$$ where %$M_e^{diag}$ is the diagonal mass matrix of charged leptons which depends on the modulus $\tau$ and the parameters $\frac{\lambda_1}{\lambda_3}$ and %$\frac{\lambda_2}{\lambda_3}$ up to overall parameter. To deal only with the left-handed mixing, it is convenient to use the Hermitian matrix $M_e=m_e^{\dagger} ~m_e$.

The neutrino mass matrix in the basis $(\nu_L, N^c, S )$ is given by
\bea
M=\left(
          \begin{array}{ccc}
            0 & m_D & 0 \\
            m_D^T & 0 & M_R \\
            0 & M_R^T & \mu_s \\
          \end{array}
        \right).
        \eea
After diagonalization of this matrix, one can get three eigenvalues, one for the light neutrino and the other two for the heavy neutrino states. The masses of the light neutrino state $m_\nu$ can be obtained as
\bea
m_{\nu}=m_D {M_R}^{-1} \mu_s {M_R^T}^{-1}m_D^T.
\eea
The overall parameter $\frac{f v_u^2 g_1^2\lambda_c^3}{h^2 v^{\prime}}$ determines the scale of light neutrino masses and can be easily chosen to achieve the desired scale. For instant, we can set $h\sim {\cal O}(1$ GeV), $f\sim {\cal O}(0.001$ GeV), $v^{\prime}\sim {\cal O}(100 $ TeV), $v_u\sim {\cal O}~(10^2$ GeV) and $g_1\sim{\cal O}(0.01$ GeV) to get the neutrino masses of order ${\cal O}(10^{-1}$ eV).  The neutrino mass matrix $m_{\nu}$ is complex and symmetric, so it is convenient to diagonalize the Hermitian matrix $M_{\nu}=m_{\nu}^{\dagger} m_{\nu}$,
\bea
M_{\nu}^{diag}=U_{\nu}^{\dagger}M_{\nu}U_{\nu} .
\eea
The lepton mixing $U_{PMNS}$ matrix is given by
\bea
U_{PMNS}=U_e^{\dagger} U_{\nu}.
\eea
The mixing angles can be calculated from the relations
\bea
Sin^2(\theta_{13})=|{(U_{PMNS})_{13}}|^2, ~Sin^2(\theta_{12})=\frac{|{(U_{PMNS})_{12}}|^2}{1-|{(U_{PMNS})_{13}}|^2}, ~Sin^2(\theta_{23})=\frac{|{(U_{PMNS})_{23}}|^2}{1-|{(U_{PMNS})_{13}}|^2}.
\eea
\begin{table}
\centering
\begin{tabular}{|c|c|c|c|c|c|c|c|c|}
  \hline
   &$\frac{\Delta m^2_{12}}{(10^{-5}~\text{eV}^2)}$ & $\frac{|\Delta m^2_{23}|}{(10^{-3}~\text{eV}^2)}$ & $r=\frac{\Delta m^2_{12}}{|\Delta m^2_{23}|}$&$\theta_{12}/^{\circ}$ & $\theta_{23}/^{\circ}$ &$ \theta_{13}/^{\circ} $&$\delta_{CP}/\pi$ \\
   \hline
  best fit&7.39&2.51&0.0294&33.82&49.8&8.6&1.57\\
   \hline
  $3\sigma$ range& 6.79-8.01 & 2.41-2.611 &0.026-0.033& 31.61-36.27 & 40.6-52.5& 8.27-9.03 &1.088-2  \\
  \hline
\end{tabular}
\caption{3$\sigma$ range for neutrino mixings and mass difference
squares from \cite{Esteban:2018azc} for inverted hierarchy.}\label{recentdata}
\end{table}
The mixing angles and mass ratios are determined by the ratios $g_2/g_1$, $\frac{\lambda_1}{\lambda_3}$, $\frac{\lambda_2}{\lambda_3}$  and the modulus $\tau$. The parameter $g_2/g_1$ is complex in general, so we can write it as
\bea
\frac{g_2}{g_1}=g e^{i \phi},
\eea
where $\phi$ is the relative phase of $g_1$ and $g_2$.
The best fit values and $3 \sigma$ ranges for the experimental results are summarized in Table (\ref{recentdata}), in which the neutrino mass squared differences are defined as
$$\Delta m_{12}^2=m_2^2-m_1^2, ~~~|\Delta m_{23}^2|=|m_3^2-(m_2^2+m_1^2)/2|.$$ %For simplicity we consider $\phi=0$.
%\textbf{case I, $\phi=0.$} \\
The parameters are scanned in the upper half of the complex plane by fixing $r=\frac{\Delta m^2_{12}}{|\Delta m^2_{23}|}$ and the mixing angles with the $3\sigma$ ranges in table (\ref{recentdata}). The modulus $\tau$ is scanned in the ranges $Re[\tau]\in[-0.5, 0.5]$ and $Im[\tau]\in[0.4, 1]$, the coupling ratio $g$ is scanned in the range $g\in[0.5,3]$ and the phase $\phi\in[-\pi, \pi]$. We study the model in the case of normal and inverted hierarchies.

  %to find the points satisfy the present experimental data of mixing angles and mass ratios shown in Table(\ref{recentdata})\cite{Esteban:2018azc}.%The correct mass ratio of the charged leptons are obtained when $\frac{\lambda_1}{\lambda_3}=41.76, ~\frac{\lambda_2}{\lambda_3}=588$
\subsection{Normal hierarchy}
For the normal hierarchy, we found the following benchmark:\\
                                       $\tau=-0.245+ 0.5236i, ~~g=2.503,~ \phi=-0.105 \pi,~~\frac{\lambda_1}{\lambda_3}=0.00031, ~\frac{\lambda_2}{\lambda_1}=0.063$, with
\bea
r &=&\frac{\Delta m^2_{12}}{|\Delta m^2_{23}|}=0.0293, ~~~~\frac{m_e}{m_{\tau}}=0.0003,~~~~\frac{m_{\mu}}{m_{\tau}}=0.061,\nonumber\\\theta_{12}&=&33.25^\circ,~~ \theta_{23}=41.678^\circ, ~~\theta_{13}=8.73^\circ.
\eea
\subsection{Inverted hierarchy}
For inverted neutrino mass hierarchy, we found the following benchmarks:\\\begin{enumerate}
                                      \item $\tau=-0.494+ 0.55i, ~~g=2.05,~~ \phi=-\pi/2,~~\frac{\lambda_1}{\lambda_3}=0.0009, ~~\frac{\lambda_2}{\lambda_1}=0.07$, with
\bea
r &=&\frac{\Delta m^2_{12}}{|\Delta m^2_{23}|}=0.0286, ~~~~\frac{m_e}{m_{\tau}}=0.0003,~~~~\frac{m_{\mu}}{m_{\tau}}=0.061,\nonumber\\\theta_{12}&=&32.4^\circ,~~ \theta_{23}=49.26^\circ, ~~\theta_{13}=8.54^\circ.
\eea
                                      \item $\tau=0.0962+0.984i, ~~g=2.05, \phi=-\pi/2,~~\frac{\lambda_1}{\lambda_3}=0.0009, ~\frac{\lambda_2}{\lambda_1}=0.07$,  with
\bea
r &=&\frac{\Delta m^2_{12}}{|\Delta m^2_{23}|}=0.0296, ~~~~\frac{m_e}{m_{\tau}}=0.0003,~~~~\frac{m_{\mu}}{m_{\tau}}=0.061,\nonumber\\\theta_{12}&=&32.36^\circ,~~ \theta_{23}=49.24^\circ, ~~\theta_{13}=8.73^\circ.
\eea
                                    \end{enumerate}

The two points $\tau=-0.494+ 0.55~i$ and $\tau=0.0962+0.984~i$ are close to the fixed points $\tau_1=-0.5+0.5i$ and $\tau_C=i$ respectively. The two fixed points are related to each other as $\tau_1=ST \tau_C$. The modular group $A_4$ is broken to its subgroup $Z_2=\{I, S \}$ at $\tau_C=i$ as neutrino and charged lepton mass matrices are invariant under the $S$ transformation \cite{Novichkov:2018ovf, Novichkov:2018yse}. The lepton masses and mixing at $\tau_C$ are studied in \cite{Novichkov:2018ovf, Novichkov:2018yse}
with the conclusion that $\tau_C$ can not be used to lead to the correct lepton masses and mixing.
 The point $\tau_1=-0.5+0.5i$ is invariant under $ST^2ST$ transformation $\tau=\frac{-(1+\tau)}{1+2\tau}$ at which the group $A_4$ is broken to its subgroup $Z_2=\{I, ST^2ST\}$. The charged lepton mass matrix $M_e$ is invariant under unitary transformation $S_1=ST^2 ST$,
$S_1^{\dagger}M_e S_1=M_e$, where
\bea
S_1=ST^2 ST=\frac{1}{3}\left(
                        \begin{array}{ccc}
                         -1 & 2\omega^2 & 2\omega \\
                        2\omega & -1 & 2\omega^2 \\
                       2\omega^2 & 2\omega & -1 \\
                    \end{array}
                 \right).\label{s1 transformation}
\eea
One of the eigenvalues of $M_e$ is zero since the determent of $M_e$ vanishes, so $\tau_1$ can not be used to lead to the correct charged lepton masses. The matrix $M_{\nu}$ is invariant under the transformation $S_1=ST^2 ST$ in Eq.(\ref{s1 transformation}). In this case, one of the eigenvalues of $M_{\nu}$ is zero and $Det(M_{\nu})=0$. The mixing matrix in this case has two vanishing mixing angles and a nearly maximal angle. As we see that the observed lepton masses and mixing are consequences of breaking modular residual symmetry by deviation from $\tau_C$ or $\tau_1$.

\section{Quark masses}
The embedding of the quark sector into a flavor model is a challenge due to the differences in the mass hierarchy and mixing for leptons and quarks. In this model, we extend the modular $A_4$ symmetry to the quark sector at the value of the modulus $\tau_1=-0.494+0.55~i$. All quarks transform as singlets under $A_4$.  The assignments of the quark fields are shown in table\ref{quarks}.

\begin{table}
  \centering
  \begin{tabular}{|c|c|c|c|c|c|c|c|c|c|}
  \hline
  fields & $Q_1$ &$Q_2$&$Q_3$ &$u_1^c$ & $u_2^c$ & $u_3^c$ & $d_1^c$ & $d_2^c$ & $d_3^c$ \\
   \hline
  $A_4$ & 1 & $1^{\prime}$ & $1^{\prime\prime}$ & $1^{\prime\prime}$&$1$&$1^{\prime}$&1&$1^{\prime\prime}$&$1^{\prime}$ \\
   \hline
  $k_I$ & 3 & 2 & 0 & 4& 4&1&0&-1&4  \\
  \hline
\end{tabular}
  \caption{Assignment of quarks under $A_4$ and the modular weight $k_I$}\label{quarks}
\end{table}

The $A_4$ invariant superpotential for down quarks can be written as
\bea
w_d=\frac{h_{11}^d}{\Lambda^3} d_1^c H_d Q_1 \chi^3+\frac{h_{22}^d}{\Lambda} d_2^c H_d Q_2 \chi+\frac{h_{23}^d}{\Lambda^2}Y_2^{(4)} d_3^c H_d Q_2 \chi^2+ h_{33}^d Y_1^{(4)} d_3^c H_d Q_3.
\eea
The chosen $A_4$ and $k_I$ assignments prevent the other mixing terms. Without loss of generality, we assume that $h_{11}^d/h_{33}^d\sim h_{22}^d/h_{33}^d\sim 1/2, $ and $h_{23}^d/h_{33}^d\sim 1$. The down quark mass matrix takes the form
\bea
m_d=h_{33}^d~ <H_d> \left(
      \begin{array}{ccc}
       \lambda^3/2 & 0 & 0 \\
        0 & \lambda/2 & 0 \\
        0 & Y_2^{(4)} \lambda^2 & Y_1^{(4)} \\
      \end{array}
    \right).
\eea
To deal with left handed mixing only, we construct the matrix $M_d=m_d^{\dagger}m_d$ which can be diagonalized by
%This mass matrix can be diagonalization by biunitary transformation, $V_d^{L{\dagger}}~m_d~V^R_d=M_d$, where
%\bea
%V^L_d&=&\left(
%\begin{array}{ccc}
% 1 & 0& 0\\
% 0 & -0.999& 0.0022 \\
% 0& 0.0022-0.00003 i & 0.999-0.014 i \\
%\end{array}
%\right),\nonumber\\~
\bea
V_d&=& \left(
\begin{array}{ccc}
 1 & 0 & 0 \\
 0& -0.490 - 0.869 i & 0.0257 + 0.045 i\\
 0& 0.0524 & 0.9985 \\
\end{array}
\right),
\eea
with the corresponding eigenvalues
\bea
M_d=diag(\lambda^4/2, \lambda^2/2, 1)~h_{33}^d~Y_1^{(4)}~<H_d>.
\eea
The hierarchical spectrum of the down quark masses is in a good agreement with the recent data for quark masses: \cite{PDG}
$$m_d=4.67_{- 0.17}^{+ 0.48}\text{MeV}, ~~~m_s=93_{-5}^{+11}\text{MeV}~~~m_b=4.18_{-0.02}^{+0.03}\text{GeV}.$$

For the up quarks, using the condition $Y_3^{(4)}=0$, the invariant superpotential under modular $A_4$ can be written as
\bea
w_u&=&\frac{h_{11}^u }{\Lambda^3}Y_2^{(4)} u_1^c H_u Q_1\chi^3+\frac{h_{12}^u }{\Lambda^2}Y_1^{(4)} u_1^c H_u Q_2\chi^2+\frac{ h_{21}^u}{\Lambda^3}Y_1^{(4)} u_2^c H_u Q_1\chi^3\nonumber\\&+&h_{23}^u Y_2^{(4)} u_2^c H_u Q_3+\frac{h_{33}^u}{\Lambda} u_3^c H_u Q_3\chi.
\eea
The up quark mass matrix takes the form
\bea
m_u= <H_u> \left(
  \begin{array}{ccc}
   h_{11}^u Y_2^{(4)} \lambda^3&h_{12}^u Y_1^{(4)} \lambda^2 & 0 \\
   h_{21}^u Y_1^{(4)} \lambda^3 & 0 &h_{23}^u Y_2^{(4)} \\
    0 & 0 &  h_{33}^u \lambda \\
  \end{array}
\right).
\eea
Assume that the couplings $h_{11}^u\sim h_{12}^u\sim h_{21}^u\sim h_{33}^u,~ h_{23}^u\sim 3 h_{33}^u$,
the Hermitian matrix $M_u=m_u^{\dagger}m_u$ is diagonalized by
\bea
V_u=\left(
\begin{array}{ccc}
 -0.478+0.848 i & -0.11+0.198 i & 0.0017\, -0.0035i \\
 -0.118-0.195 i & 0.504\, +0.83 i & 1.5\times 10^{-7} \\
 0.00349 & 0.0008245
   & 0.999942 \\
\end{array}
\right),
\eea
with the corresponding eigenvalues
\bea
M_u^{diag}=h_{33}^u <H_u> Y_1^{(4)} diag(\lambda^7, \lambda^3, 1),
\eea
which are in agreement with the up quark mass ratios \cite{PDG}
$$\frac{m_u}{m_t}=0.000012, ~\frac{m_c}{m_t}=0.008.$$ The quark mixing matrix, $V_{CKM}$ takes the form
\bea
|V_{CKM}|= |V_u^{\dagger}~V_d|=
\left(
\begin{array}{ccc}
 0.9737 & 0.224 & 0.006 \\
 0.224 & 0.9723 & 0.05 \\
 0.005 & 0.05 & 0.9986 \\
\end{array}
\right),
\eea
which is close to the correct $V_{CKM}$ \cite{PDG}. The same result of quark masses and mixing can be obtained at $\tau=0.0964+0.984i$ while the observed masses and mixing are not achieved
at $\tau=0.2525+0.526i$ with the above quark model.
\section{Conclusion}
We built an $A_4$ modular invariance model to account for both lepton and quark masses and mixing. The model is free from large number of flavons or extra symmetries like $Z_N$ symmetries which were considered in many models based on the flavor symmetry. The neutrino masses are obtained via inverse seesaw mechanism. The predicted lepton mixing and mass ratios are compatible with the recent data. The neutrino mass square difference ratios and lepton mixing angles are determined in terms of the coupling ratio $g_2/g_1$ and the modulus $\tau$ at values near fixed points for inverted hierarchy scenario.  For the same value of $\tau=-0.494+0.55~i$, we extend the modular $A_4$ symmetry to the quark sector. The calculated quark mass ratios and mixing are in a quite agreement with the experimental results.


\begin{thebibliography}{99}%%%%%%%%%%
%%%%%%%%%%%%%%%%%%%%%%%%%%%%%%%%%%%%%%%%%%%%%%%%%%%%%%%
%\cite{Wyler:1982dd}
\bibitem{T1} P.~Minkowski,
  %``$\mu \to e\gamma$ at a Rate of One Out of $10^{9}$ Muon Decays?,''
  Phys.\ Lett.\ B {\bf 67},(1977) 421; R.~N.~Mohapatra and G.~Senjanovic,
  %``Neutrino Mass and Spontaneous Parity Violation,''
  Phys.\ Rev.\ Lett.\  \textbf{44}, 912 (1980); T.~Yanagida,
 % ``Horizontal Symmetry and Masses of Neutrinos,''
  Prog.\ Theor.\ Phys.\  \textbf{64}, 1103 (1980);
M.~Gell-Mann, P.~Ramond and R.~Slansky,
%``Complex Spinors and Unified Theories,''
Conf. Proc. C \textbf{790927}, 315-321 (1979)
[arXiv:1306.4669 [hep-th]]



\bibitem{Wyler:1982dd}
  D.~Wyler and L.~Wolfenstein,
 % ``Massless Neutrinos in Left-Right Symmetric Models,''
  Nucl.\ Phys.\ B \textbf{218}, 205 (1983).
  %%CITATION = NUPHA,B218,205;%%
  %257 citations counted in INSPIRE as of 03 Aug 2015

%\cite{Mohapatra:1986bd}
\bibitem{Mohapatra:1986bd}
  R.~N.~Mohapatra and J.~W.~F.~Valle,
%  ``Neutrino Mass and Baryon Number Nonconservation in Superstring Models,''
  Phys.\ Rev.\ D {\bf 34}, 1642 (1986).
  %%CITATION = PHRVA,D34,1642;%%
  %654 citations counted in INSPIRE as of 03 ao�t 2015

%\cite{Ma:1987zm}
\bibitem{Ma:1987zm}
  E.~Ma,
 % ``Lepton Number Nonconservation in $E(6)$ Superstring Models,''
  Phys.\ Lett.\ B {\bf 191}, 287 (1987).
  %%CITATION = PHLTA,B191,287;%%
  %33 citations counted in INSPIRE as of 03 ao�t 2015

 \bibitem{'t Hooft}
G. t Hooft, Phys. Rev. Lett. {\bf 37}, 8 (1976);

 \bibitem{discrete-symmetries}
  A selective list includes: W.~Grimus, A.~S.~Joshipura,
S.~Kaneko, L.~Lavoura and M.~Tanimoto,
%  ``Lepton mixing angle theta(13) = 0 with a horizontal symmetry D(4),''
  JHEP {\bf 0407}, 078 (2004)
  [arXiv:hep-ph/0407112];
J.~Kubo, A.~Mondragon, M.~Mondragon and E.~Rodriguez-Jauregui,
 % ``The flavor symmetry,''
  Prog.\ Theor.\ Phys.\  {\bf 109}, 795 (2003)
  [Erratum-ibid.\  {\bf 114}, 287 (2005)]
  [arXiv:hep-ph/0302196];
R.~N.~Mohapatra, M.~K.~Parida and G.~Rajasekaran,
 % ``High scale mixing unification and large neutrino mixing angles,''
  Phys.\ Rev.\  D {\bf 69}, 053007 (2004)
  [arXiv:hep-ph/0301234];
C.~Hagedorn, M.~Lindner and R.~N.~Mohapatra,
 % ``S(4) flavor symmetry and fermion masses: Towards a grand unified theory  of
  flavor,''
  JHEP {\bf 0606}, 042 (2006)
  [arXiv:hep-ph/0602244];
I.~de Medeiros Varzielas, S.~F.~King and G.~G.~Ross,
 % ``Neutrino tri-bi-maximal mixing from a non-Abelian discrete family
 %ymmetry,''
  Phys.\ Lett.\  B {\bf 648},201 (2007)
  [arXiv:hep-ph/0607045];
E.~Ma,
%  ``Near tri-bimaximal Neutrino Mixing with Delta(27) Symmetry,''
  Phys.\ Lett.\  B {\bf 660}, 505 (2008)
  [arXiv:0709.0507 [hep-ph]];
C.~Luhn, S.~Nasri and P.~Ramond,
 % ``Tri-Bimaximal Neutrino Mixing and the Family Symmetry $Z_7 \times Z_3$,''
  Phys.\ Lett.\  B {\bf 652},) 27 (2007
  [arXiv:0706.2341 [hep-ph]];
E.~Ma and G.~Rajasekaran,
 % ``Softly broken A(4) symmetry for nearly degenerate neutrino masses,''
  Phys.\ Rev.\  D {\bf 64}, 113012 (2001)
  [arXiv:hep-ph/0106291];


 \bibitem{modulargroups}
 %\cite{deAdelhartToorop:2011re}
  R.~de Adelhart Toorop, F.~Feruglio and C.~Hagedorn,
 % ``Finite Modular Groups and Lepton Mixing,''
  Nucl.\ Phys.\ B {\bf 858}, 437 (2012)
  doi:10.1016/j.nuclphysb.2012.01.017
  [arXiv:1112.1340 [hep-ph]].
  %%CITATION = doi:10.1016/j.nuclphysb.2012.01.017;%%
  %162 citations counted in INSPIRE as of 01 Dec 2019

  %\cite{Feruglio:2017spp}
\bibitem{Feruglio:2017spp}
  F.~Feruglio,
%  ``Are neutrino masses modular forms?,''
  doi:$10.1142/9789813238053_0012$
  arXiv:1706.08749 [hep-ph].

%\cite{Kobayashi:2018vbk}
\bibitem{Kobayashi:2018vbk}
  T.~Kobayashi, K.~Tanaka and T.~H.~Tatsuishi,
 % ``Neutrino mixing from finite modular groups,''
  Phys.\ Rev.\ D {\bf 98} no.1,  016004 (2018)
  doi:10.1103/PhysRevD.98.016004
  [arXiv:1803.10391 [hep-ph]].
  %%CITATION = doi:10.1103/PhysRevD.98.016004;%%
  %42 citations counted in INSPIRE as of 02 Dec 2019
%\cite{Kobayashi:2019rzp}

%\cite{Okada:2019xqk}
\bibitem{Okada:2019xqk}
  H.~Okada and Y.~Orikasa,
 % ``A modular $S_3$ symmetric radiative seesaw model,''
  arXiv:1907.04716 [hep-ph].
  %%CITATION = ARXIV:1907.04716;%%
  %13 citations counted in INSPIRE as of 02 Dec 2019

%\cite{Du:2020ylx}
\bibitem{Du:2020ylx}
X.~Du and F.~Wang,
%``SUSY Breaking Constraints on Modular flavor $S_3$ Invariant $SU(5)$ GUT Model,''
[arXiv:2012.01397 [hep-ph]].
%1 citations counted in INSPIRE as of 22 Jan 2021

%\cite{Xing:2019edp}
\bibitem{Xing:2019edp}
Z.~Z.~Xing and D.~Zhang,
%``Seesaw mirroring between light and heavy Majorana neutrinos with the help of the S$_{3}$ reflection symmetry,''
JHEP \textbf{03}, 184 (2019)
doi:10.1007/JHEP03(2019)184
[arXiv:1901.07912 [hep-ph]].
%5 citations counted in INSPIRE as of 22 Jan 2021

%\cite{Kobayashi:2018scp}
\bibitem{Kobayashi:2018scp}
  T.~Kobayashi, N.~Omoto, Y.~Shimizu, K.~Takagi, M.~Tanimoto and T.~H.~Tatsuishi,
%  ``Modular A$_{4}$ invariance and neutrino mixing,''
  JHEP {\bf 1811}, 196(2018),
  doi:10.1007/JHEP11196 (2018)
  [arXiv:1808.03012 [hep-ph]].
  %%CITATION = doi:10.1007/JHEP11(2018)196;%%
  %38 citations counted in INSPIRE as of 02 Dec 2019

%\cite{Okada:2018yrn}
\bibitem{Okada:2018yrn}
  H.~Okada and M.~Tanimoto,
 % ``CP violation of quarks in $A_4$ modular invariance,''
  Phys.\ Lett.\ B {\bf 791} 54(2019),
  doi:10.1016/j.physletb.2019.02.028
  [arXiv:1812.09677 [hep-ph]].
  %%CITATION = doi:10.1016/j.physletb.2019.02.028;%%
  %27 citations counted in INSPIRE as of 02 Dec 2019

%\cite{Ding:2019zxk}
\bibitem{Ding:2019zxk}
G.~J.~Ding, S.~F.~King and X.~G.~Liu,
%``Modular A$_{4}$ symmetry models of neutrinos and charged leptons,''
JHEP \textbf{09}, 074(2019),
doi:10.1007/JHEP09(2019)074
[arXiv:1907.11714 [hep-ph]].
%28 citations counted in INSPIRE as of 04 May 2020

%\cite{Gui-JunDing:2019wap}
\bibitem{Gui-JunDing:2019wap}
G.~J.~Ding, S.~F.~King, X.~G.~Liu and J.~N.~Lu,
%``Modular S$_{4}$ and A$_{4}$ symmetries and their fixed points: new predictive examples of lepton mixing,''
JHEP \textbf{12}, 030(2019),
doi:10.1007/JHEP12(2019)030
[arXiv:1910.03460 [hep-ph]].
%18 citations counted in INSPIRE as of 04 May 2020

%\cite{Nomura:2019yft}
\bibitem{Nomura:2019yft}
  T.~Nomura and H.~Okada,
  %``A two loop induced neutrino mass model with modular $A_4$ symmetry,''
  arXiv:1906.03927 [hep-ph].
  %%CITATION = ARXIV:1906.03927;%%
  %15 citations counted in INSPIRE as of 02 Dec 2019

%\cite{Nomura:2019xsb}
\bibitem{Nomura:2019xsb}
  T.~Nomura, H.~Okada and S.~Patra,
 % ``An Inverse Seesaw model with $A_4$-modular symmetry,''
  arXiv:1912.00379 [hep-ph].
  %%CITATION = ARXIV:1912.00379;%%
  %7 citations counted in INSPIRE as of 04 May 2020

%\cite{Asaka:2019vev}
\bibitem{Asaka:2019vev}
T.~Asaka, Y.~Heo, T.~H.~Tatsuishi and T.~Yoshida,
%``Modular $A_4$ invariance and leptogenesis,''
JHEP \textbf{01}, 144 (2020)
doi:10.1007/JHEP01(2020)144
[arXiv:1909.06520 [hep-ph]].
%31 citations counted in INSPIRE as of 22 Jan 2021

   %\cite{Penedo:2018nmg}
\bibitem{Penedo:2018nmg}
  J.~T.~Penedo and S.~T.~Petcov,
 % ``Lepton Masses and Mixing from Modular $S_4$ Symmetry,''
  Nucl.\ Phys.\ B {\bf 939} 292 (2019)
  doi:10.1016/j.nuclphysb.2018.12.016
  [arXiv:1806.11040 [hep-ph]].
  %%CITATION = doi:10.1016/j.nuclphysb.2018.12.016;%%
  %39 citations counted in INSPIRE as of 02 Dec 2019

%\cite{deMedeirosVarzielas:2019cyj}
%\bibitem{deMedeirosVarzielas:2019cyj}
%I.~de Medeiros Varzielas, S.~F.~King and Y.~L.~Zhou,
%``Multiple modular symmetries as the origin of flavor,''
%Phys. Rev. D \textbf{101}, no.5, 055033 (2020)
%doi:10.1103/PhysRevD.101.055033
%[arXiv:1906.02208 [hep-ph]].
%47 citations counted in INSPIRE as of 19 Sep 2020

%\cite{Kobayashi:2019mna}
\bibitem{Kobayashi:2019mna}
  T.~Kobayashi, Y.~Shimizu, K.~Takagi, M.~Tanimoto and T.~H.~Tatsuishi,
 % ``New $A_4$ lepton flavor model from $S_4$ modular symmetry,''
  arXiv:1907.09141 [hep-ph].
  %%CITATION = ARXIV:1907.09141;%%
  %15 citations counted in INSPIRE as of 02 Dec 2019

 %\cite{Okada:2019lzv}
\bibitem{Okada:2019lzv}
H.~Okada and Y.~Orikasa,
%``Neutrino mass model with a modular $S_4$ symmetry,''
[arXiv:1908.08409 [hep-ph]].
%23 citations counted in INSPIRE as of 22 Jan 2021

%\cite{Kobayashi:2019xvz}
\bibitem{Kobayashi:2019xvz}
T.~Kobayashi, Y.~Shimizu, K.~Takagi, M.~Tanimoto and T.~H.~Tatsuishi,
%``$A_4$ lepton flavor model and modulus stabilization from $S_4$ modular symmetry,''
Phys. Rev. D \textbf{100}, no.11, 115045 (2019)
[erratum: Phys. Rev. D \textbf{101}, no.3, 039904 (2020)]
doi:10.1103/PhysRevD.100.115045
[arXiv:1909.05139 [hep-ph]].
%41 citations counted in INSPIRE as of 22 Jan 2021

%\cite{Wang:2019ovr}
\bibitem{Wang:2019ovr}
X.~Wang and S.~Zhou,
%``The minimal seesaw model with a modular S$_{4}$ symmetry,''
JHEP \textbf{05}, 017 (2020)
doi:10.1007/JHEP05(2020)017
[arXiv:1910.09473 [hep-ph]].
%33 citations counted in INSPIRE as of 23 Jan 2021

%\cite{Novichkov:2018nkm}
\bibitem{Novichkov:2018nkm}
  P.~P.~Novichkov, J.~T.~Penedo, S.~T.~Petcov and A.~V.~Titov,
 % ``Modular A$_{5}$ symmetry for flavour model building,''
  JHEP {\bf 1904} 174 (2019)
  doi:10.1007/JHEP04(2019)174
  [arXiv:1812.02158 [hep-ph]].
  %%CITATION = doi:10.1007/JHEP04(2019)174;%%
  %31 citations counted in INSPIRE as of 02 Dec 2019

%\cite{Ding:2019xna}
\bibitem{Ding:2019xna}
  G.~J.~Ding, S.~F.~King and X.~G.~Liu,
 % ``Neutrino Mass and Mixing with $A_5$ Modular Symmetry,''
  arXiv:1903.12588 [hep-ph].
  %%CITATION = ARXIV:1903.12588;%%
  %26 citations counted in INSPIRE as of 02 Dec 2019

 %\cite{Criado:2019tzk}
\bibitem{Criado:2019tzk}
J.~C.~Criado, F.~Feruglio and S.~J.~D.~King,
%``Modular Invariant Models of Lepton Masses at Levels 4 and 5,''
JHEP \textbf{02}, 001 (2020)
doi:10.1007/JHEP02(2020)001
[arXiv:1908.11867 [hep-ph]].
%36 citations counted in INSPIRE as of 22 Jan 2021

%\cite{Okada:2020rjb}
\bibitem{Okada:2020rjb}
H.~Okada and M.~Tanimoto,
%``Quark and lepton flavors with common modulus $\tau$ in $A_4$ modular symmetry,''
[arXiv:2005.00775 [hep-ph]].
%18 citations counted in INSPIRE as of 26 Dec 2020

%\cite{Okada:2019uoy}
\bibitem{Okada:2019uoy}
H.~Okada and M.~Tanimoto,
%``Towards unification of quark and lepton flavors in $A_4$ modular invariance,''
[arXiv:1905.13421 [hep-ph]].
%48 citations counted in INSPIRE as of 26 Dec 2020

%\cite{Liu:2020akv}
\bibitem{Liu:2020akv}
X.~G.~Liu, C.~Y.~Yao and G.~J.~Ding,
%``Modular Invariant Quark and Lepton Models in Double Covering of $S_4$ Modular Group,''
[arXiv:2006.10722 [hep-ph]].
%18 citations counted in INSPIRE as of 26 Dec 2020

%\cite{Kobayashi:2019rzp}
\bibitem{Kobayashi:2019rzp}
T.~Kobayashi, Y.~Shimizu, K.~Takagi, M.~Tanimoto and T.~H.~Tatsuishi,
%``Modular $S_3$-invariant flavor model in SU(5) grand unified theory,''
PTEP \textbf{2020}, no.5, 053B05 (2020)
doi:10.1093/ptep/ptaa055
[arXiv:1906.10341 [hep-ph]].
%48 citations counted in INSPIRE as of 26 Dec 2020

%\cite{Kobayashi:2018wkl}
\bibitem{Kobayashi:2018wkl}
T.~Kobayashi, Y.~Shimizu, K.~Takagi, M.~Tanimoto, T.~H.~Tatsuishi and H.~Uchida,
%``Finite modular subgroups for fermion mass matrices and baryon/lepton number violation,''
Phys. Lett. B \textbf{794}, 114-121 (2019)
doi:10.1016/j.physletb.2019.05.034
[arXiv:1812.11072 [hep-ph]].
%73 citations counted in INSPIRE as of 26 Dec 2020

%\cite{Yao:2020qyy}
\bibitem{Yao:2020qyy}
C.~Y.~Yao, J.~N.~Lu and G.~J.~Ding,
%``Modular Invariant $A_{4}$ Models for Quarks and Leptons with Generalized CP Symmetry,''
[arXiv:2012.13390 [hep-ph]].
%0 citations counted in INSPIRE as of 28 Dec 2020

%\cite{Lu:2019vgm}
\bibitem{Lu:2019vgm}
J.~N.~Lu, X.~G.~Liu and G.~J.~Ding,
%``Modular symmetry origin of texture zeros and quark lepton unification,''
Phys. Rev. D \textbf{101}, no.11, 115020 (2020)
doi:10.1103/PhysRevD.101.115020
[arXiv:1912.07573 [hep-ph]].
%25 citations counted in INSPIRE as of 22 Jan 2021

%\cite{Novichkov:2018yse}
\bibitem{Novichkov:2018yse}
P.~P.~Novichkov, S.~T.~Petcov and M.~Tanimoto,
%``Trimaximal Neutrino Mixing from Modular A4 Invariance with Residual Symmetries,''
Phys. Lett. B \textbf{793}, 247-258 (2019)
doi:10.1016/j.physletb.2019.04.043
[arXiv:1812.11289 [hep-ph]].
%59 citations counted in INSPIRE as of 19 Sep 2020

%\cite{Novichkov:2018ovf}
\bibitem{Novichkov:2018ovf}
% ``Modular S$_{4}$ models of lepton masses and mixing,''
  JHEP {\bf 1904} 005(2019)
  doi:10.1007/JHEP04(2019)005
  [arXiv:1811.04933 [hep-ph]].
  %%CITATION = doi:10.1007/JHEP04(2019)005;%%
  %34 citations counted in INSPIRE as of 02 Dec 2019

%\cite{deMedeirosVarzielas:2019cyj}
\bibitem{deMedeirosVarzielas:2019cyj}
I.~de Medeiros Varzielas, S.~F.~King and Y.~L.~Zhou,
%``Multiple modular symmetries as the origin of flavor,''
Phys. Rev. D \textbf{101}, no.5, 055033 (2020)
doi:10.1103/PhysRevD.101.055033
[arXiv:1906.02208 [hep-ph]].
%58 citations counted in INSPIRE as of 28 Dec 2020

%\cite{Yao:2020zml}
\bibitem{Yao:2020zml}
C.~Y.~Yao, X.~G.~Liu and G.~J.~Ding,
%``Fermion Masses and Mixing from Double Cover and Metaplectic Cover of $A_5$ Modular Group,''
[arXiv:2011.03501 [hep-ph]].
%5 citations counted in INSPIRE as of 28 Dec 2020

%\cite{Wang:2020lxk}
\bibitem{Wang:2020lxk}
X.~Wang, B.~Yu and S.~Zhou,
%``Double Covering of the Modular $A^{}_5$ Group and Lepton Flavor Mixing in the Minimal Seesaw Model,''
[arXiv:2010.10159 [hep-ph]].
%6 citations counted in INSPIRE as of 28 Dec 2020

%\cite{Novichkov:2020eep}
\bibitem{Novichkov:2020eep}
P.~P.~Novichkov, J.~T.~Penedo and S.~T.~Petcov,
%``Double Cover of Modular $S_4$ for Flavour Model Building,''
[arXiv:2006.03058 [hep-ph]].
%18 citations counted in INSPIRE as of 28 Dec 2020

%\cite{Liu:2019khw}
\bibitem{Liu:2019khw}
X.~G.~Liu and G.~J.~Ding,
%``Neutrino Masses and Mixing from Double Covering of Finite Modular Groups,''
JHEP \textbf{08}, 134 (2019)
doi:10.1007/JHEP08(2019)134
[arXiv:1907.01488 [hep-ph]].
%57 citations counted in INSPIRE as of 22 Jan 2021

%\cite{Okada:2020brs}
\bibitem{Okada:2020brs}
H.~Okada and M.~Tanimoto,
%``Spontaneous CP violation by modulus $\tau$ in $A_4$ model of lepton flavors,''
[arXiv:2012.01688 [hep-ph]].
%2 citations counted in INSPIRE as of 28 Dec 2020

%\cite{Novichkov:2019sqv}
\bibitem{Novichkov:2019sqv}
P.~P.~Novichkov, J.~T.~Penedo, S.~T.~Petcov and A.~V.~Titov,
%``Generalised CP Symmetry in Modular-Invariant Models of Flavour,''
JHEP \textbf{07}, 165 (2019)
doi:10.1007/JHEP07(2019)165
[arXiv:1905.11970 [hep-ph]].
%57 citations counted in INSPIRE as of 28 Dec 2020

%\cite{Kobayashi:2019uyt}
\bibitem{Kobayashi:2019uyt}
T.~Kobayashi, Y.~Shimizu, K.~Takagi, M.~Tanimoto, T.~H.~Tatsuishi and H.~Uchida,
%``$CP$ violation in modular invariant flavor models,''
Phys. Rev. D \textbf{101}, no.5, 055046 (2020)
doi:10.1103/PhysRevD.101.055046
[arXiv:1910.11553 [hep-ph]].
%27 citations counted in INSPIRE as of 28 Dec 2020

%\cite{Nomura:2020cog}
\bibitem{Nomura:2020cog}
T.~Nomura and H.~Okada,
%``Modular $A_4$ symmetric inverse seesaw model with $SU(2)_L$ multiplet fields,''
[arXiv:2007.15459 [hep-ph]].
%5 citations counted in INSPIRE as of 23 Jan 2021





\bibitem{Bruinier2008The}
J.~H. Bruinier, G.~V.~D. Geer, G.~Harder, and D.~Zagier, {\em The 1-2-3 of
  Modular Forms}.
\newblock Universitext. Springer Berlin Heidelberg, 2008.

\bibitem{diamond2005first}
F.~Diamond and J.~M. Shurman, {\em A first course in modular forms}, vol.~228
  of {\em Graduate Texts in Mathematics}.
\newblock Springer, 2005.

\bibitem{Gunning1962}
R.~C. Gunning, {\em Lectures on Modular Forms}.
\newblock Princeton, New Jersey USA, Princeton University Press, 1962.


\bibitem{Esteban:2018azc}
  I.~Esteban, M.~C.~Gonzalez-Garcia, A.~Hernandez-Cabezudo, M.~Maltoni and T.~Schwetz,
 % ``Global analysis of three-flavour neutrino oscillations: synergies and tensions in the determination of $\theta_{23}$, $\delta_{CP}$, and the mass ordering,''
  JHEP {\bf 1901}, 106 (2019)
  doi:10.1007/JHEP01(2019)106
  [arXiv:1811.05487 [hep-ph]].
  %%CITATION = doi:10.1007/JHEP01(2019)106;%%
  %218 citations counted in INSPIRE as of 23 Nov 2019

\bibitem{PDG}
M. Tanabashi et al. (Particle Data Group), Phys. Rev. D 98, 030001 (2018) and 2019 update.
\end{thebibliography}
\end{document}